\documentclass[journal=jacsat,manuscript=article]{achemso}
\usepackage[version=3]{mhchem} 
\usepackage{tikz}

\usepackage{amsmath,amssymb}
\usepackage{bm}

\usepackage[symbol]{footmisc}

\author{Fadhila Chehami}
\email{fadhila.chehami@unilim.fr}
\author{Cyril Decroze}
\affiliation{University of Limoges, XLIM, UMR 7252, F-87000 Limoges, France}
\author{Thomas Pasquet}
\author{Emmanuel Perrin}
\affiliation{Centre d’Ingénierie des Systèmes en Télécommunications en ElectroMagnétisme et Electronique (CISTEME), F-87000 Limoges, France}
\author{Thomas Fromenteze}
\affiliation{University of Limoges, XLIM, UMR 7252, F-87000 Limoges, France}
\title[An \textsf{achemso} demo]
  {Morphogenetic Design of Self-Organized Correlated Disordered \textcolor{black}{Electromagnetic Media}}

\abbreviations{IR,NMR,UV}

\begin{document}

\begin{abstract}
The last decades witnessed the emergence of the field of correlated disordered media, a great challenge offering a large panel of new perspectives for applications in theoretical modelling and material fabrication. The efficient design of structures with a controlled level of spatial correlation is a central challenge in this field, in a context where existing techniques generally rely on gradient descent on non-convex functions and on the use of stochastic methods to explore vast design spaces more efficiently. In this work, we propose a new generative technique based on Alan Turing's morphogenesis theory for designing correlated disordered materials. Inspired by the structuring of living organisms, this technique relies on the definition of simple local interactions guiding the self-organization of a generated medium. The decentralization of design constraints and the elimination of cost function minimization make this approach natively scalable to the design of large domains with controlled levels of disorder. As a validation, the morphogenetic generation of stealthy hyperuniform disordered structures is exploited to reproduce an experiment of isotropic electromagnetic bandgap synthesis in the microwave range using a low refractive index contrast. 
  
\end{abstract}

\section{Keywords}
Morphogenesis, self-organized pattern, stealthy hyperuniform disordered structure, isotropic photonic band gap

\section{Introduction}

The design of novel materials with unprecedented properties is a central challenge of modern science. Indeed, in order to achieve a better control of the electromagnetic wave propagation, the scientific community has proposed several designed structured media~\cite{koenderink2015nanophotonics}. 

Among the latter, physicists distinguish the photonic crystals, defined as artificial materials with periodically arranged elements, directly influencing the light generation and propagation~\cite{joannopoulos1997photonic}. The most prominent property observed, in this kind of arrangements, is the exhibition of large frequency ranges over which electromagnetic wave propagation is prohibited, the so-called photonic/electromagnetic band gaps~\cite{soukoulis2012photonic}. Since their introduction in 1987~\cite{yablonovitch1987inhibited}, the photonic band gaps in periodic structures, have given rise to various promising applications such as telecommunication devices~\cite{larsen2003optical}, sensors~\cite{zhang2015review} and efficient radiation sources~\cite{thevenot1999directive,chang2006efficient}. However, because of their highly symmetric distributions, the devices based on photonic crystals have several shortcomings such as anisotropic properties, strictly linked with the eigenaxes of the crystal, and high sensitivity to defects and manufacturing dispersions.

Following the discovery of the absence of diffusion of electrons in correlated disordered media\footnote[1]{in the framework of P. Anderson's research awarded with the Nobel Prize in physics 1977~\cite{anderson1958absence}}, this new class of materials have emerged and gained growing attention~\cite{yu2021engineered}. At the interface between crystalline and perfectly uncorrelated patterns, these structures benefit from properties deriving from both order and disorder and their study has given rise to an active field of research combining wave physics in complex media and nanophotonics. In this context, several works exploited the inherent structural correlations of partially disordered media in order to control the scattering~\cite{florescu2013optical, horodynski2022anti, milovsevic2019hyperuniform}, transport~\cite{leseur2016high, davy2021experimental} and localization of light in matter~\cite{ricouvier2019foam, lin2020chip}. 

\textcolor{black}{This trend is motivated by the many disordered photonic materials found in nature that reveal intriguing structural color effects in plants, insects, and mammals~\cite{sun2013structural}. Indeed, over the last decade, numerous research works succeeded in demonstrating that the exploitation of the structuring of certain biological media, spontaneously occurring in nature, could be an optimized solution allowing to reach efficient functionalities in different fields of science. For instance, in ~\cite{barry2020evolutionary} the most stereotypical natural photonic structures were retrieved, such as Bragg mirrors, chirped dielectric mirrors or the gratings on top of Morpho-butterfy wings, using evolutionary algorithms inspired by natural evolution. Moreover, photonic band gaps in self-assemblies of amorphous gyroids inspired by the wing scale structuring in the butterfly \textit{Pseudilycaena} have been demonstrated~\cite{sellers2017local}. Narasimhan \textit{et al.} synthetized transparent photonic nanostructures inspired by the longtail glasswing butterfly \textit{Chorinea faunus} exhibiting a good angle-independent white-light transmission, strong hydrophilicity and anti biofouling properties, which prevent adhesion of proteins, bacteria and eukaryotic cells~\cite{narasimhan2018multifunctional}. The study in~\cite{moyroud2017disorder} revealed that the effective degree of order/disorder in floral structuring generates a photonic signature that is highly salient to insect pollinators. Furthermore, the highest white retroreflection of the white beetles has been tightly associated with the disordered networks constituting their wings~\cite{wilts2018evolutionary}. Also, in~\cite{plamann2020relating} the authors linked the transparency in some living tissues such as the eye cornea to the organization of the collagen protein constituting the latter.}

This work is also inspired by the research of R. Doursat, who pioneered the conceptualization of morphogenetic engineering aimed at developing decentralized design methods. By imagining the adaptation of biological mechanisms to the emergence of macroscopic functions from small-scale self-organization, he and his co-authors have thus proposed entirely new perspectives in computer science, robotics, electricity, mechanics and civil engineering~\cite{doursat2013review}. 

Over the years, several studies proposed different numerical approaches to generate structures with a controlled level of order/disorder~\cite{molesky2018inverse, vynck2021light}. The first design protocol which succeeded to create a strongly correlated media was introduced by Florescu \textit{et al.}~\cite{florescu2009designer} using a gradient descent algorithm optimizing over two degrees of freedom. Besides, some disordered media were generated by the random packing of hard spheres in different dimensions (using the random sequential absorption model~\cite{widom1966random}, the Lubachevsky-Stillinger algorithm~\cite{lubachevsky1990geometric} or the ideal amorphous solids approach~\cite{lee2010geometry}). Furthermore, other studies exploited two-body interaction potentials such as Lennard-Jones together with standard Monte-Carlo techniques~\cite{de2016self} and the molecular dynamics based on various softwares such as NAMD~\cite{phillips2005scalable} and CHARMMS~\cite{brooks2009charmm}. Nevertheless, almost all the above mentioned techniques rely on gradient descent algorithms for the minimization of functions over many degrees of freedom, converging towards local minima. Thus, their adaptation to large scale problems or to specific conditions turn out to be a daunting computational task despite the recent development of the optimization methods~\cite{rechtsman2008optimized}. \textcolor{black}{Most recently, inverse design methods based on Metropolitan selection rule minimizing cost functions in terms of the autocovariance have been exploited to generate two-dimensional non-Hermitian disordered lattices~\cite{piao2022wave}. Another study exploited evolving scattering networks for engineering disorder in time-varying networks based on the minimization of cost functions describing the probabilistic preferential attachment of new particles to more connected ones~\cite{yu2023evolving}.}

 In this paper, we propose a new design method of correlated disordered structures inspired by the morphogenesis theory introduced by Alan Turing, in an attempt to explain the spontaneous formation of patterns in nature~\cite{turing1990chemical}. Exploiting the mechanisms of reaction and diffusion of chemical species, the Gray-Scott model~\cite{gray1983autocatalytic} is adapted in this work as a generative technique. This mathematical model is particularly simple to implement, easily adaptable to complex geometries and highly parallelizable for application to large domains. These properties are provided by the definition of simple laws of interaction at a local scale, decentralizing the design constraints and guiding the structuring of media controlled by a number of reaction-diffusion parameters. Although this design protocol can be used to create different types of self-organized disordered media in two or more dimensions~\cite{Sims}, we focus in this paper on its first adaptation to the design self-organized correlated disordered structures displaying large isotropic electromagnetic band gaps.

Our paper is organized as follows: In Section 1, we describe the morphogenetic generation method used to design our correlated disordered media. In Section 2, we present the numerical and experimental results carried out on the morphogenetic structures, namely the band diagram simulations and the transmission measurements over all incident angles, underscoring the exhibition of large isotropic photonic band gaps in the microwave range. Section 3 provides concluding remarks.    

\section{Morphogenetic generation method}
\begin{figure}
\includegraphics[width=0.5\columnwidth]{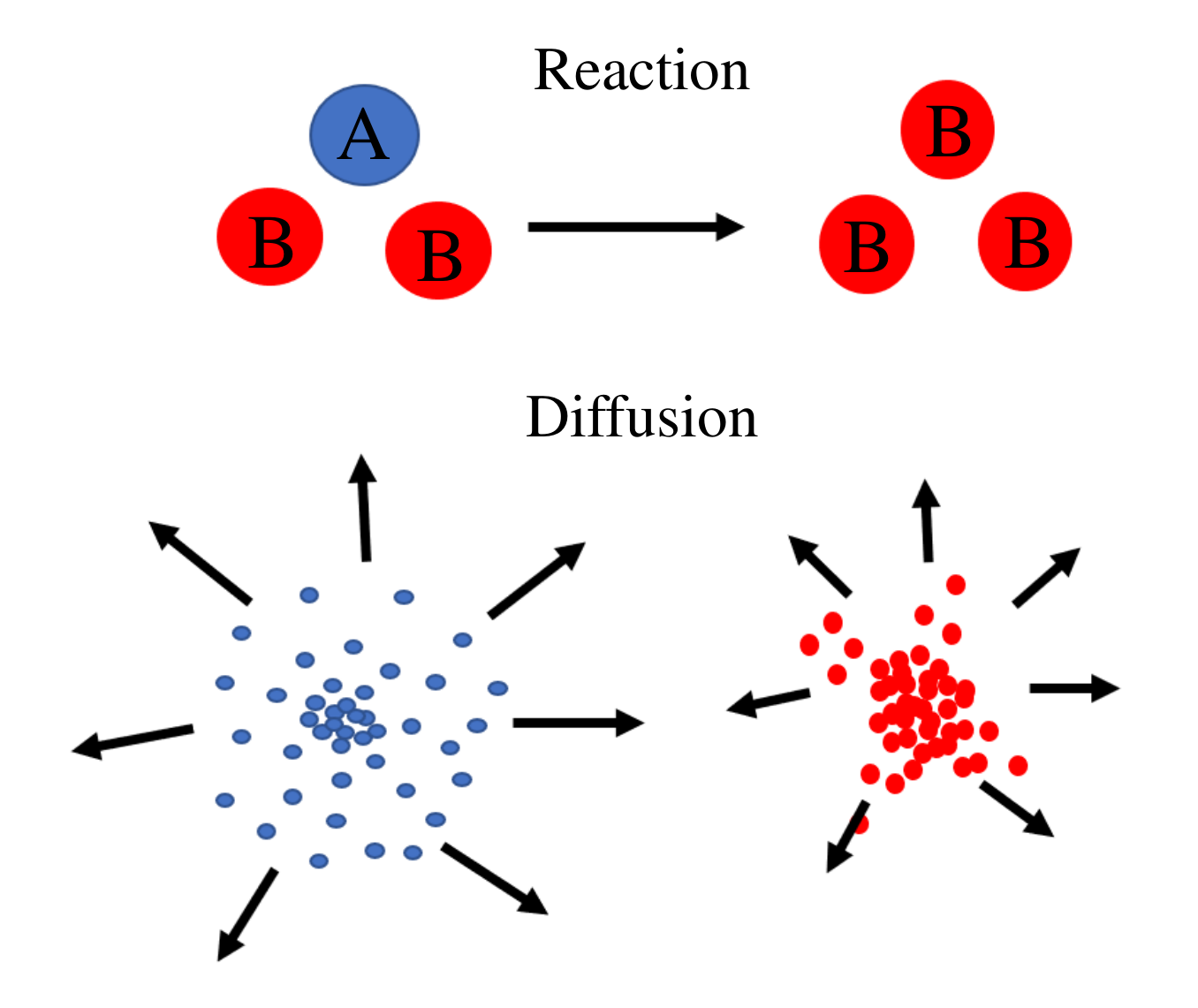}
\caption{Reaction-diffusion phenomena occurring between two chemical species $A$ and $B$. The diffusion, conversion and regulation of the morphogens lead to population equilibria associated with the emergence of spatial patterns.\label{fig1}}
\end{figure}

Nature, in its large biodiversity exhibits a canvas of self-organized patterns of fascinating geometrical forms and colors. Through the centuries, the research field tried to study, explain and be inspired by this behaviour. A. Turing was the first one to propose a reaction-diffusion system in an attempt to explain the spontaneous formation of self-regulated patterns in nature~\cite{turing1990chemical}. The generative system proposed by Gray and Scott is undoubtedly the most popular reaction-diffusion model~\cite{gray1983autocatalytic}. It is based on a simple system of two chemical species interacting with each other and generating various spatial patterns autonomously~\cite{kondo2010reaction}. The reaction-diffusion between the two chemical species $A$ and $B$ (referred to as morphogens by A. Turing~\cite{turing1990chemical}), illustrated in Fig.~\ref{fig1}, is governed by the following partial differential equations: 
\begin{align}
\frac{\partial \mathbf{A}}{\partial t} = d_A \, \nabla^2 \, \mathbf{A} - \mathbf{A} \mathbf{B}^2 + f \, (1 - \mathbf{A}) \nonumber\\
\frac{\partial \mathbf{B}}{\partial t} = d_B \, \nabla^2 \, \mathbf{B} + \mathbf{A} \mathbf{B}^2 - (f + k) \, \mathbf{B} 
\label{eq:one}
\end{align}

These equations are solved using the Euler method resulting in the equation system below, determining the concentration of the chemical species through a succession of discrete time iterations of a finite difference model:
\begin{align}
\mathbf{A}_{n+1} = \mathbf{A}_n + (d_A\nabla^2 \mathbf{A}_n - \mathbf{A}_n\mathbf{B}_n^2 + f\,(1-\mathbf{A}_n))\, \Delta t \nonumber \\
\mathbf{B}_{n+1} = \mathbf{B}_n + (d_B\nabla^2 \mathbf{B}_n + \mathbf{A}_n\mathbf{B}_n^2 - (k+f)\,\mathbf{B}_n)\, \Delta t 
\label{eq:two}
\end{align}

$\mathbf{A}$ and $\mathbf{B}$ are arrays composed of respective morphogen concentrations (unitless and ranging from $0$ to $1$). This model thus corresponds to a simple chemical reaction computed at each time iteration $n$. On the one hand, it describes the reaction phenomena converting an element $A$ in $B$ when in the presence of two $B$. On the other hand, the species $A$ and $B$ diffuse in space according to coefficients $d_A$ and $d_B$ under the action of a Laplacian operator $\nabla^2$. Population equilibrium between the two species is ensured with appropriate set of parameters of spontaneous generation of the $A$ particles, defined as $f$ (feed rate), and spontaneous extinction of the $B$ particles, defined as $k$ (kill rate).

\begin{figure}
\includegraphics[width=0.55\columnwidth]{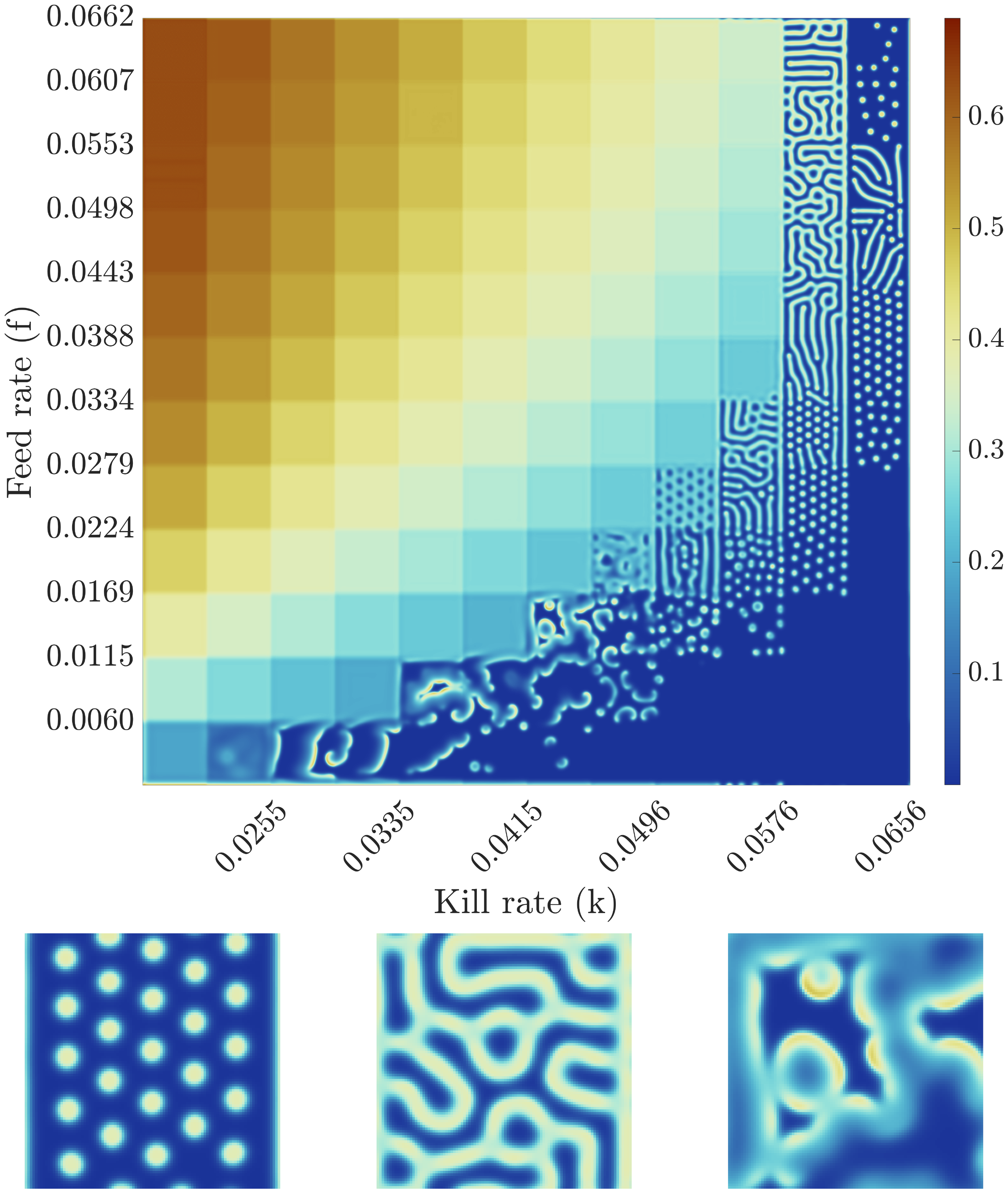}
\caption{Distribution of the chemical species $A$ concentration for different sets of parameters $k$ and $f$ with $d_A=0.5$, $d_B=1$. The bottom row shows the zoom of some typical generated patterns,in the GS model, namely (from the left to the right): spots, worms and the Belousov-Zhabotinsky pattern~\cite{zhabotinsky2007belousov}.\label{fig2}}
\end{figure}

Depending on the sets of parameters and the initial conditions applied, the reaction-diffusion model gives rise to a rich and complex pattern panel (Fig.~\ref{fig2}) studied and classified by Pearson~\cite{pearson1993complex}. The bottom row of Fig.~\ref{fig2}, depicts, in particular, self-replicating spots, worms and the so called Belousov-Zhabotinsky pattern~\cite{zhabotinsky2007belousov}. 

\begin{figure}[h!]
\includegraphics[width=0.55\columnwidth]{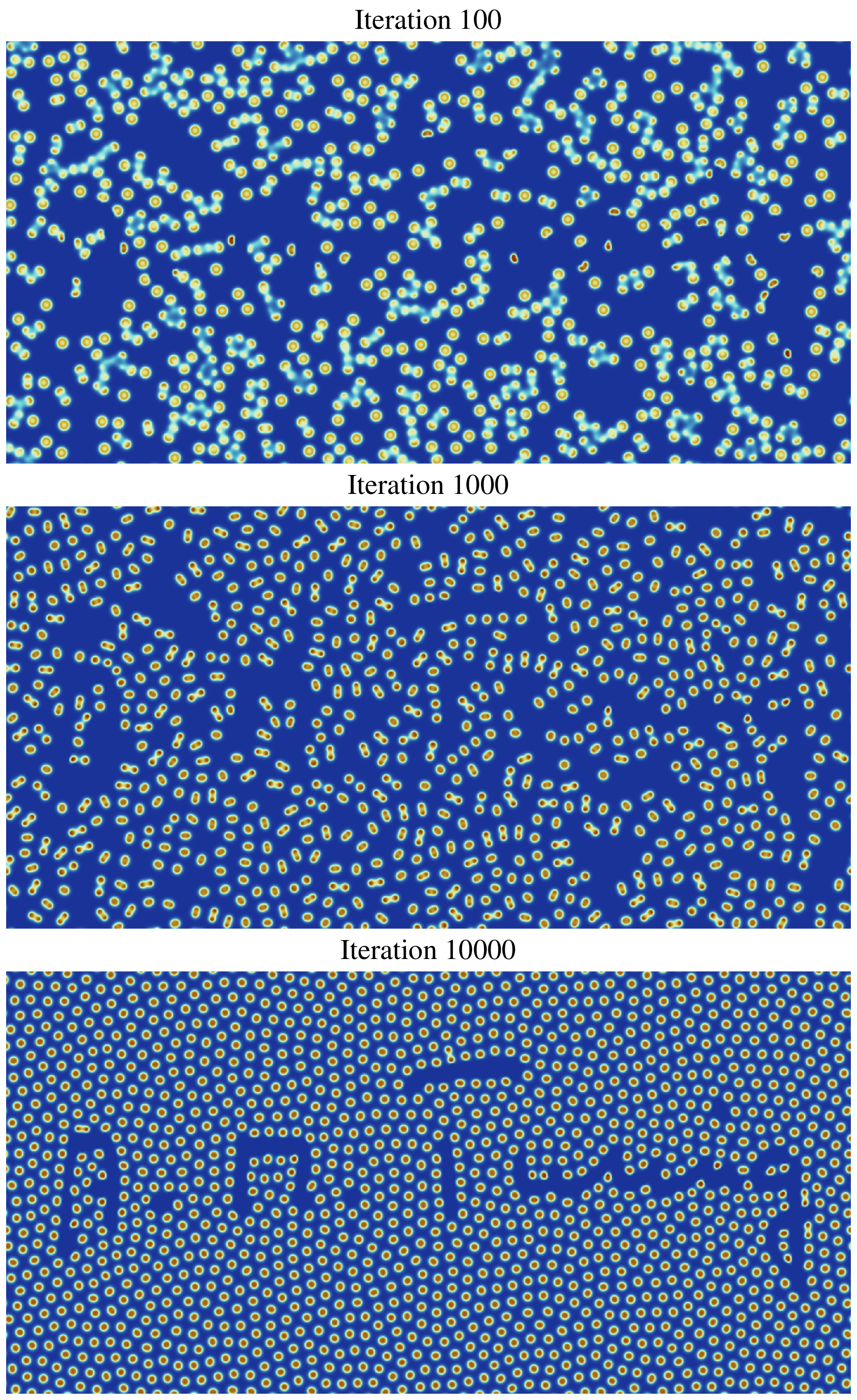}
\caption{Synthesis of a constrained structure of self-replicating spots with $f=0.036 $, $k= 0.065$, $d_A=0.5$, $d_B=1$, at different time iterations when a shape forming Alan Turing's signature is set as a prohibited region, forcing the concentration of A morphogens to 0 at the corresponding locations.\label{fig3}}
\end{figure}

In this work, we focus on the specific class of self-replicating spots generated with $f=0.036 $, $k= 0.065$, $d_A=0.5$, and $d_B=1$, for the procedural generation of materials exhibiting electromagnetic band gaps. To illustrate the synthesis of such patterns and show their intrinsic property of adaptation to geometrical constraints, Fig.~\ref{fig3} depicts the spatial distributions of A morphogens through a succession of time iterations of the generative model. In this case, after setting some initial spots at random locations in the simulated domain, the growth of pattern is forbidden in an area corresponding to Alan Turing's signature. The initial spots undergo a self-replicating behavior according to a mechanism similar to mitosis. The process continues until the entire simulation domain is covered with circular spots, fitting the imposed conditions and tending to fill the available space with a densely packed structure.   

This property of self-organization corresponds to natural characteristics of Turing patterns to optimize the distances between particles, transiting from pseudo-random arrangements to compact crystals. This technique replaces in this work the minimization of interaction potentials between particles through computationally intensive gradient descent. The following section presents a study of the synthesized disorder properties and an adaptation of this technique to the generation of isotropic band gaps.

\section{Results and discussions}
\subsection{Disorder control and generation of electromagnetic band gaps}

This study aims at demonstrating that the self-organization capacity of Turing patterns during a reaction-diffusion synthesis allows a direct control of the disorder level of the generated media. The exploitation of these properties will ensure the formation of isotropic electromagnetic band gaps, thus enabling the reproduction of the experimental results of Florescu \textit{et al.}~\cite{man2013photonic}. Ensuring the synthesis of morphogenetic correlated disordered media with the presented technique, the formation of isotropic band gaps in the microwave region represents an illustration of one of the multiple properties observed in such structures. \textcolor{black}{Since the frequency range of the observed band gaps is strictly related to the dimensions of the self-replicating spots and their arrangements, which in turn are controlled by the parameters of the Gray-Scott model, band diagram computations using the conventional plane wave expansion method~\cite{johnson2001block, rumpf2015engineering} were performed in order to optimize the latter and ensure band gap formation in the microwave region. The development of the plane wave expansion method used in these studies is available in the supporting information.}

At this point, it is worth mentioning that the existence of band gaps in correlated disordered structures is counter intuitive since the photonic crystals studies stipulate that the formation of a band gap requires Bragg scattering and long-range translational order, properties lacking in this kind of structures. Hence, several research works tried to propose an explanation of the physical mechanisms behind this intriguing phenomena~\cite{froufe2016role, froufe2017band}. Among the latter, the most popular one found in the literature, inspired by the tight-binding model in solid-state physics, states that the coupled resonances between short-range correlated neighboring scatterers is responsible in the opening of a band gap in such structures with a correlated disorder.

\textcolor{black}{The generative model described by Eq.~(\ref{eq:two}) is thus implemented in a domain of 300 by 300 pixels with $f=0.036 $, $k= 0.065$, $d_A=0.5$, $d_B=1$ resulting in circular spots spaced by an average distance of $5$\,mm . This process is then ensured during the desired number of temporal iterations depending on the order level of the synthesized patterns for which metrics will be introduced later. Then, following a thresholding of the synthesized patterns, the continuous concentrations of the morphogens are converted into a biphasic simulation domain. A $0.45$ thresholding parameter was selected resulting in circular spots with an average diameter of $3$\,mm. After that, a relative permittivity $\epsilon_r = 3$ is assigned to the circular spots on a background material defined by $\epsilon_r = 1$. These values are considered to conceive an experimental demonstration of a microwave photonic band gap based on the exploitation of acrylic cylinders arranged in air. It is interesting to note that among the favorable properties offered by this generative model, the implementation of the finite difference Laplacian operator ensures that periodic boundary conditions are satisfied. The synthesis of periodic supercells is thus directly achieved by the model, facilitating the simulation of infinite media for the calculation of band diagrams.}

In addition to the dispersion diagram computations, the study of the properties of the structures synthesized with this morphogenetic technique requires the evaluation of the obtained degree of order/disorder. The translational order metric~$\tau$, first introduced by Torquato \textit{et al.}~\cite{distasio2018rational}, is adapted in this work. This order metric is defined as a measure of the statistical deviation of a given partially correlated media compared to a reference uncorrelated disorder following a Poisson distribution. The patterns are discretized in binary pixels, having an area $N_s = L_1 \times L_2$, $L_1$ and $L_2$ are the side lengths. They are mathematically represented by a function $\sigma(m,n)$ which takes two integers as input ($m$ and $n$, the indices specifying the pixel location in the lattice) and yields a binary output (either $0$ or $1$ for unoccupied and occupied sites respectively). The order metric $\tau$ is calculated using the following formula:  
\begin{align}
    \tau = \frac{1}{\tau_{max}} \sum _{k\neq0} \,[\,\mathbf{S(k)} - \mathbf{S_p(k)}\,]\,^2 = \frac{1}{\tau_{max}} \sum _{k\neq0}\, [\,\mathbf{S}(\mathbf{k}) - (1-f_f)\,]\,^2
    \label{eq:three}
\end{align}

\noindent where the filling fraction $f_f=N/N_s$ corresponds to the ratio of occupied sites $N$ over the total number of pixels $N_s$ and $\mathbf{S}(\mathbf{k})$ is the static structure factor calculated over the wave vectors $\mathbf{k}$:    
\begin{align}
\mathbf{S(k)} = \frac{1}{N}\, |\,\bm{\rho}(\mathbf{k})\,|\,^2 = \frac{1}{N}\, \left|\,\sum_{m=1}^{L_1}\sum_{n=1}^{L_2} \sigma(m,n)\,e\,^{i(k_xm+k_yn)}\,\right|^2  
\label{eq:four}
\end{align}

Explicitly, the order metric is here normalized by $\tau_{max}$, corresponding to the maximum order metric computed for a checkerboard pattern of identical dimensions. The latter, alternating at pixel size between occupied and unoccupied sites with $f_f = 0.5$ is found to be the most crystalline arrangement that can be obtained for a matrix of given dimensions~\cite{distasio2018rational}.  \textcolor{black}{It is important to note that the next developments are based on homogenization conditions of the studied media, allowing the extraction of statistical properties useful for our applications. The synthesis of super-cells with natively periodic boundary conditions facilitates again the satisfaction of these assumptions.}

\begin{figure*}
\includegraphics[width=1\columnwidth]{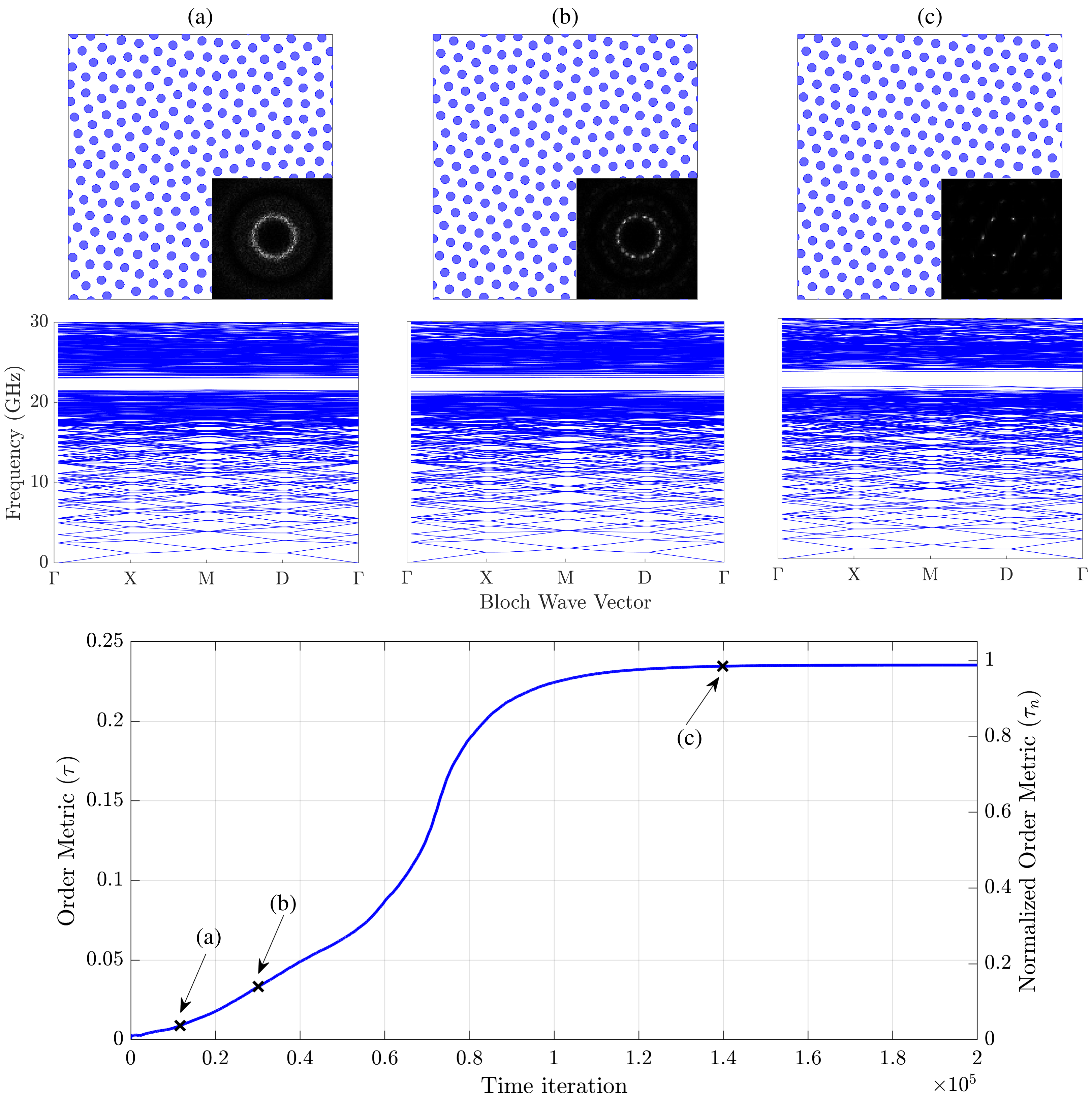}
\caption{(\textit{Top row}) Spatial distribution of the self-replicating spots in a $300\times300$ pixels domain through time iterations of the morphogenetic model, revealing three distinct states: Correlated disordered structure (a), transitional state (b) and hexagonal crystal (c). The structure factor corresponding to each distribution is exhibited in the insets. (\textit{Middle row}) Band diagram computed for each state. (\textit{Bottom row}) Variation of the order metric ($\tau$) and the normalized one ($\tau_n$) through the time iterations.\label{fig4}}
\end{figure*}

By means of the elements previously introduced, it is finally possible to compare the geometrical characteristics of the correlated disorder media generated by this new approach with the associated electromagnetic properties (Fig.~\ref{fig4}). 

The self-replicating spots goes through different self-organized arrangements with the iteration of the generative model, from a correlated disordered distribution to a crystalline one passing through a transitional case in between (Fig.~\ref{fig4}. a, c and b respectively). An animation is available in the supporting information, generated on a $300\times300$ pixel grid in $100$ seconds on a computer equipped with a Core i9-10900K processor. \textcolor{black}{The spatial distributions of the simulated supercells are compared based on their respective structure factors representing a useful information about the electromagnetic behavior of a medium, defining the main directions of diffraction in the framework of a single diffusion approximation}. As expected, the initial distribution exhibits a continuous scattering ring with no obvious directional preference corresponding to the so called stealthy hyperuniform disordered structure~\cite{torquato2015ensemble}. The last distribution obtained, after convergence of the synthesis, exhibits six distinct Bragg scattering peaks corresponding to the six fold rotational symmetry of an hexagonal crystal. The formation of weakly perturbed crystals within the structure at the transition between these two extreme distributions is finally associated with an intermediate structure factor, revealing first local symmetries.

\textcolor{black}{Fig.~\ref{fig4} depicts Transverse Magnetic band diagrams corresponding to each distribution (namely, the correlated disordered structure, the transitional one and the hexagonal crystal). All of the three states exhibit TM band gaps, in the microwave range, with an average $\Delta \omega /\omega_c = 6 \%$, $\Delta \omega /\omega_c = 7 \%$ and $\Delta \omega /\omega_c = 8 \%$ respectively ($\Delta \omega$ is the band gap width and $\omega_c$ is the band gap center frequency. It is worth noting that as the distribution converges towards a crystalline arrangement, the band gap width increases. Similar behavior, associated with the emergence of a constant exclusion zone between the scatterers, has been observed in~\cite{florescu2009designer}.}

\textcolor{black}{The order metric $\tau$ is computed during the morphogenetic synthesis of this structured medium until convergence of the obtained pattern. The curve plotted at the bottom of Fig.~\ref{fig4} reveals the increasing of the order metric through time iterations, indicating the development of short-range positional order, until a critical iteration ($It_c = 140000$, ensuring the following equality :  $\tau(It_c+1) - \tau(It_c) = 0$) around which it keeps a constant value ($\tau = 0.23$) when the pattern develops long-range correlation (crystallization).}

It is interesting to note that despite the convergence of the structure towards a compact hexagonal arrangement, the order metric remains relatively low insofar as each elementary pattern remains composed of a set of pixels with identical values limiting the correlation at smaller scales. It is thus proposed to normalize this metric by the highest value obtained for an hexagonal arrangement of associated dimensions. The distributions presented in Fig.~\ref{fig7} thus correspond to normalized order metrics of $\tau_n = 0.03$, $\tau_n = 0.13$, and $\tau_n = 0.98$, allowing to illustrate more clearly the transition from a weakly correlated state to a crystalline arrangement.

\begin{figure}[h!]
\includegraphics[width=0.65\columnwidth]{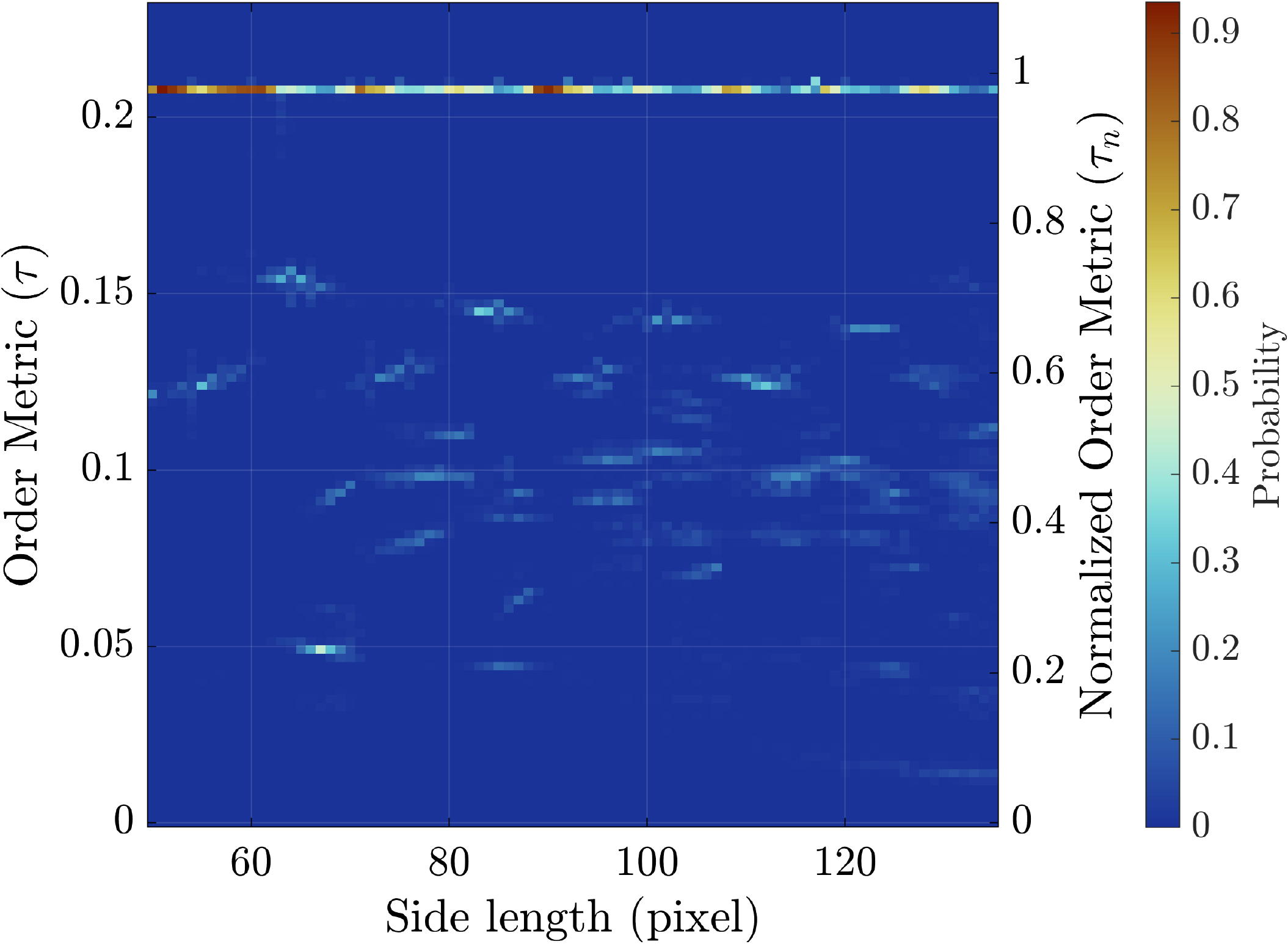}
\caption{Order metrics of morphogenetic distributions of self-replicating spots for different side lengths ($L_1 = L_2$). For each length, 200 simulations were performed to estimate the recurrence of the convergence state. \label{fig5}}
\end{figure}

It is also worth noting that the synthesis of the patterns is directly impacted by the pressured periodic boundary conditions of the domain in which the growth is simulated. We have indeed observed that, in some cases, the generated patterns no longer reach the compact hexagonal state but converge to weakly perturbed crystals. To explore this phenomenon, series of 200 simulations have been performed for different domain dimensions (Fig.~\ref{fig5}). In each case, the value of the normalized order metric $\tau_n$ is extracted only once a convergence is reached.  

\textcolor{black}{This study confirms that the convergence state varies with the domain dimensions passing from a compact crystalline arrangements to weakly perturbed crystals. This seems to be attributable to the ratio between the periods of the synthesized patterns and the domain size. Also, one can see that when increasing the domain dimensions, the number of stable states increases as well as an overall decrease in their order metrics. Indeed, the expansion of the available space allow the formation of more stable arrangements of local crystal assemblies. Moreover, Fig. \ref{fig5} reveals that the order of the most compact arrangements does not depend on the dimensions of the growth domain.}

This section demonstrated the ability of the proposed morphogenetic technique to generate correlated disordered media while offering a direct control over the disorder level. Band diagram simulations and structure factor computations were used to demonstrate the formation of electromagnetic band gaps and the synthesis of isotropic properties. 

With the help of all these elements, an experimental demonstration of this work is finally proposed in the following section.

\subsection{Band gap measurements in correlated disordered media}

The objective of this demonstration is the reproduction of an isotropic electromagnetic band gap experiment, based on the design of an arrangement of dielectric rods forming a partially correlated disorder~\cite{man2013photonic}. Replacing the optimization techniques previously used for the synthesis of such media, the proposed approach thus facilitates the generation of structures with electromagnetic properties invariant with the incident angle. 

\begin{figure}[h!]
\includegraphics[width=0.8\columnwidth]{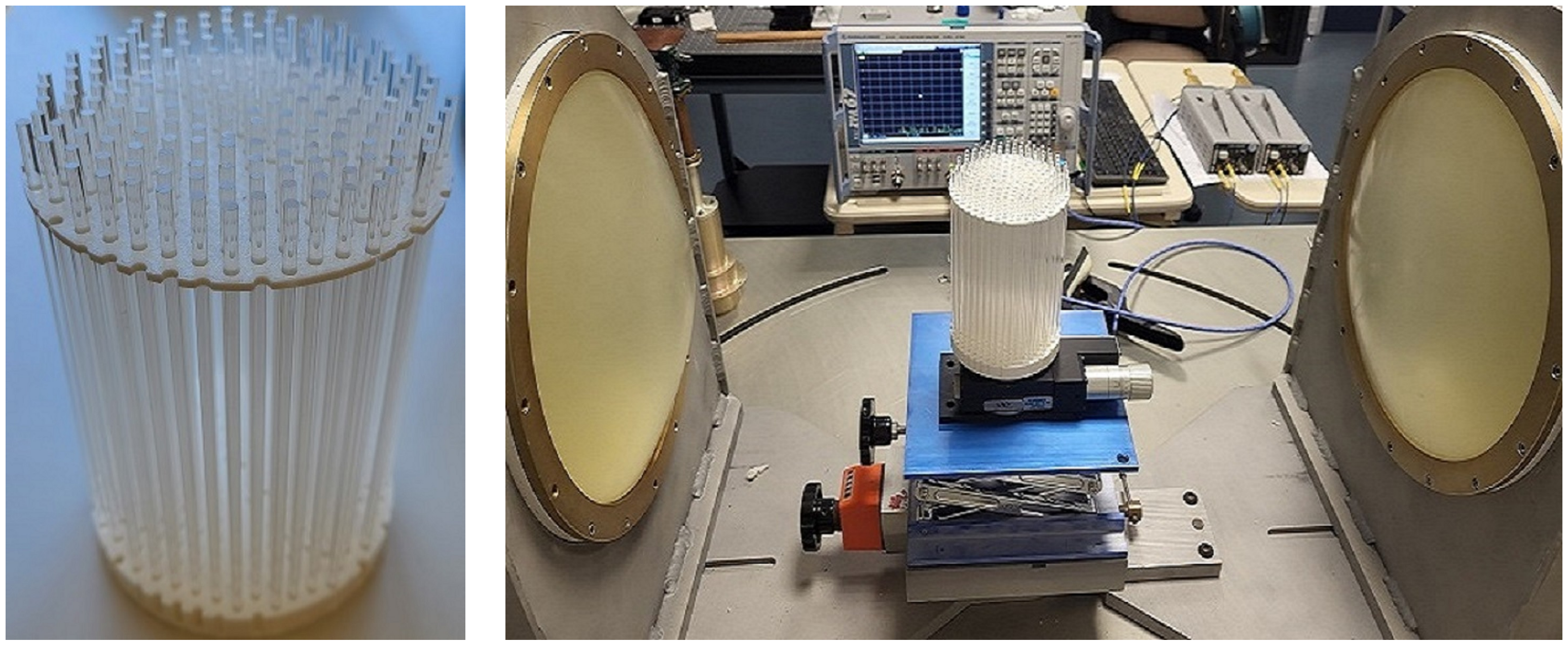}
\caption{Overview of the experimental setup used for the transmission measurements through our structures over all incident angles. The left part depicts one of the 3D samples built from commercial lollipop acrylic sticks inserted in 3D printed supports. \label{fig6}}
\end{figure}

Here, the two-dimensional morphogenetic patterns presented previously (Fig.~\ref{fig4}) were extruded along the third spatial dimension resulting in acrylic cylinders with a diameter of 3$\,$mm. In order to meet the experimental requirements, circular arrangements were designed, having a diameter of 10$\,$cm and a height of 15$\,$cm. To facilitate the set-up of this experiment and its reproducibility, commercial acrylic lollipop sticks were exploited, inserted into thin PLA-printed holders, ensuring the proper spatial distribution of the simulated elements (Fig.~\ref{fig6}).

\begin{figure}[h!]
\includegraphics[width=1\columnwidth]{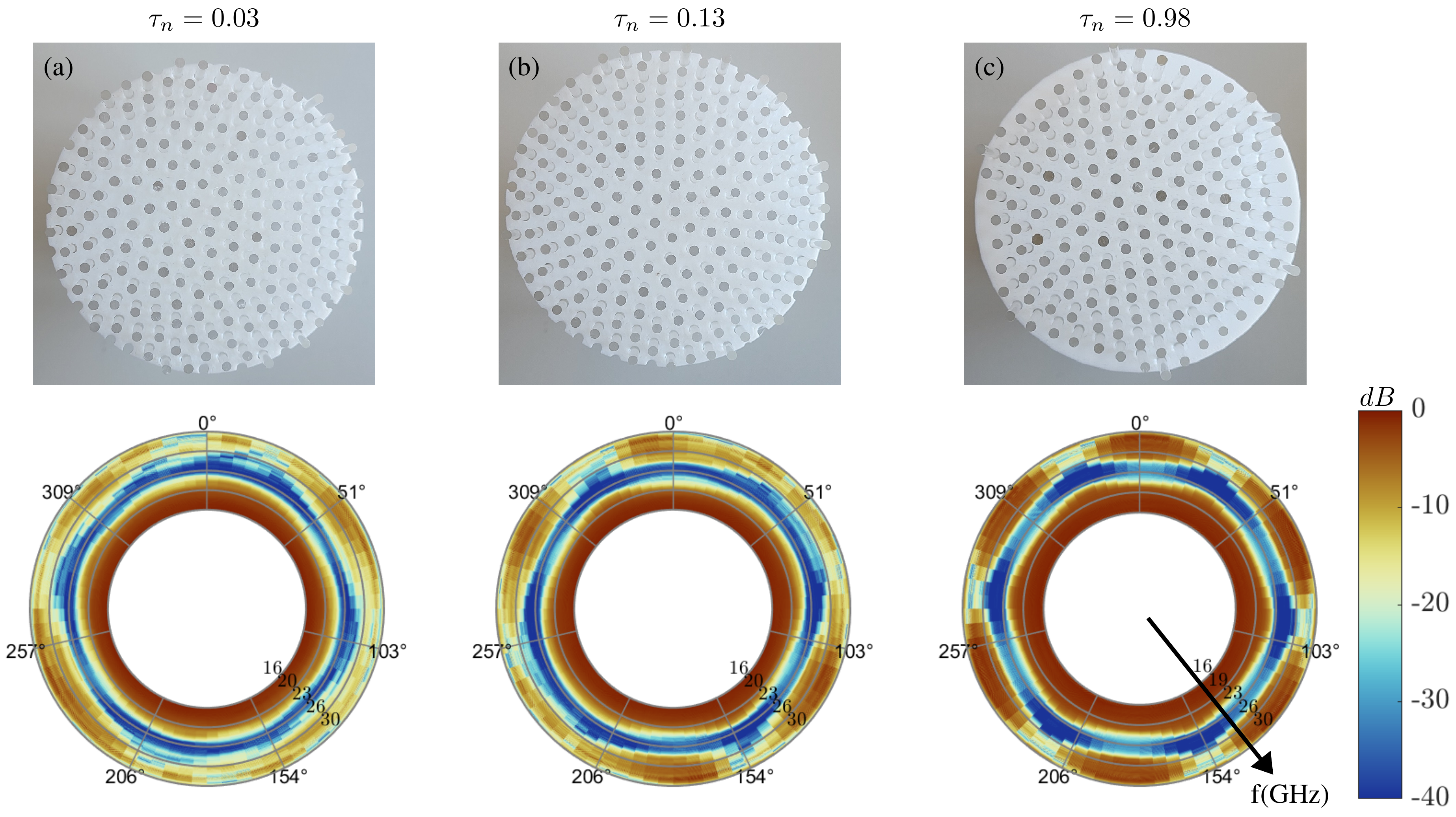}
\caption{(\textit{Top row}) Top view of the 3D morphogenetic samples: Correlated disordered structure (a), transitional state (b) and hexagonal crystal(c). (\textit{Bottom row}) TM transmission measurements over all incident angles corresponding to each sample. The transmission is plotted in polar coordinates where the radial axis corresponds to the frequency (in GHz) and the angular one corresponds to the incident angle.  \label{fig7}}
\end{figure}

Then, each sample was placed between two facing microwave horn antennas connected to a vector network analyzer (Fig.~\ref{fig6}). The device is based on lenses allowing the generation of a Gaussian beam. At the center of this beam, a locally transverse magnetic plane wave is obtained in a diameter of about $10$\,cm. Restricting the waist of this beam to the center of the formed acrylic structures thus allowed to limit possible edge effects and to ignore the contribution of the supports. Each sample was rotated about its vertical axis and the transmission was measured every five degrees. Fig.~\ref{fig7} depicts the measured transmissions as a function of frequency and incident angles in polar coordinates. The polar coordinates highlight the direct relationship between the band gap formation and the Bragg scattering planes.

These experimental results can be directly associated with the simulations presented in Fig.~\ref{fig4}. As expected, the correlated disordered structure exhibits a circular region of low transmission, common to all incidence angles, a signature of an almost isotropic electromagnetic band gap. In the opposite case of the hexagonal lattice, the frequencies of the low transmission regions depend more directly on the angle of incidence under the influence of the crystal symmetries. As expected, the transitional case exhibits a low transmission region in between the circular one and the transmission lines, where it is possible to identify the formation of eigenaxes and the emergence of anisotropic properties.
 
These results are fully consistent with those obtained from correlated disordered structures generated using gradient descent algorithms minimizing a cost function over two degrees of freedom~\cite{man2013photonic}. This replication thus attests to the effectiveness of the proposed morphogenetic design method. By proposing a generation technique where the design constraint is decentralized at a local scale through simple interaction mechanisms, we facilitate the synthesis of self-structured partially disordered media, while maintaining a high level of control over the electromagnetic properties and the imposed boundary conditions.

\section{Conclusion}
A new generation technique of correlated disordered media has been introduced, based on Turing's morphogenesis theory. In this paper, we exploited this method for the design of self-organized correlated disordered materials displaying isotropic band gaps in the microwave range. Band diagram calculations, using the plane wave expansion method, were used to design structures with electromagnetic band gaps in the desired frequency range. The natural tendency of Turing patterns to converge to compact arrangements, following the iteration of reaction diffusion models, has been exploited here to provide control over the level of order/disorder within the generated structures. The exploitation of order metrics and structure factors has allowed to highlight the rapid emergence of anisotropic properties, from the first formation of hexagonal arrangements. Replication of an experimental demonstration of isotropic bandgap using this new technique provided results in direct agreement with those published in the scientific literature. Despite the apparent complexity of the model studied, this generation technique is particularly simple to implement through the iteration of a finite difference model that can be written in only a few lines of code. This approach relies on the definition of a few local variables guiding the emergence, duplication and self-organization of compact patterns. By avoiding the exploitation of gradient descents, evaluated at a global scale on a large number of parameters, the proposed technique has the advantage of being easily scalable to high dimensional domains and easily parallelizable by the local nature of the computed interactions. \textcolor{black}{It seems finally important to note that we have restricted these prospective studies to circular patterns among all the diversity of those generable by reaction-diffusion. Future studies might highlight new properties associated with more exotic patterns (worm structures, Belousov-Zhabotinsky pattern ...) for the procedural generation of electromagnetic properties for a wide range of applications, from radio frequencies to optics, and extending to several other disciplines, such as acoustics and mechanics.}\\

\textcolor{black}{Numerous observations have identified intricate biological structures that directly influence the reflection, transmission, and absorption of light. With the fairly recent uncovering of reaction-diffusion mechanisms playing a key role in multiple facets of living organism formation, it becomes increasingly fascinating to contemplate the possibility that the generative techniques investigated in our research may also be harnessed by nature itself, meticulously orchestrating light-matter interactions following the simple definition of local chemical interactions guiding the emergence of complex functions.}

\begin{suppinfo}
\textcolor{black}{A section describing the implementation of the plane wave expansion method exploited to compute the dispersion diagrams of the morphogenetic structures.}\\
\textcolor{black}{A GIF animation illustrating the evolution of the self-replicating spots distribution, the structure factor and the density of states through the time iterations of the morphogenetic generation method. The displayed values are normalized by their maximum value.} 
\end{suppinfo}

\subsection{Funding Sources}
The authors acknowledge the support of the ANR JCJC MetaMorph (ANR-21-CE42-0005) and the Région Nouvelle-Aquitaine as part of the MorphoGem project.

\bibliography{achemso-demo}
%

\end{document}